\documentclass[12pt,preprint]{aastex}
\usepackage{natbib} 
\newcommand{\msun}{\mbox{$M_\odot$}}
\newcommand{\lsun}{\mbox{$L_\odot$}}
\newcommand{\rsun}{\mbox{$R_\odot$}}
\newcommand{\tsun}{\mbox{$T_\odot$}}
\newcommand{\teff}{\mbox{$T_{\rm eff}$}}
\newcommand{\logg}{\mbox{$\log g$}}
\newcommand{\fe}{\mbox{[Fe/H]}}

\newcommand{\etal}{\mbox{\rm et al.~}}
\newcommand{\ms}{\mbox{m s$^{-1}$}}
\newcommand{\ks}{\mbox{km s$^{-1}$}}

\newcommand{\mjup}{$M_{\rm Jup}$}
\newcommand{\mearth}{$M_{\oplus}$}

\newcommand{\rjup}{$R_{\rm Jup}$}
\newcommand{\msini}{$M \sin i~$}
\newcommand{\vsini}{$v \sin i~$}

\newcommand{\chisq}{$\sqrt{\chi_{\nu}^2}$}

\newcommand{\hipp}{$Hipparcos$}

\received{}
\accepted{}

\slugcomment{To appear in ApJ}

\shortauthors{Sato \etal}
\shorttitle{N2K}
\begin{document}

\title{The N2K Consortium. II.\\ A Transiting Hot Saturn Around HD~149026 With a Large Dense Core \altaffilmark{1,2}}
\author{Bun'ei Sato\altaffilmark{3,4},
Debra A. Fischer\altaffilmark{5}, 
Gregory W. Henry\altaffilmark{6},
Greg Laughlin\altaffilmark{7},
R. Paul Butler\altaffilmark{8},
Geoffrey W. Marcy\altaffilmark{9},
Steven S. Vogt\altaffilmark{7},
Peter Bodenheimer\altaffilmark{7},
Shigeru Ida\altaffilmark{10},
Eri Toyota\altaffilmark{3},
Aaron Wolf\altaffilmark{7},
Jeff A. Valenti\altaffilmark{11},
Louis J. Boyd\altaffilmark{12},
John A. Johnson\altaffilmark{9},
Jason T. Wright\altaffilmark{9},
Mark Ammons\altaffilmark{7},
Sarah Robinson\altaffilmark{7},
Jay Strader\altaffilmark{7},
Chris McCarthy\altaffilmark{5},
K. L. Tah\altaffilmark{5},
Dante Minniti\altaffilmark{13}}

\email{satobn@oao.nao.ac.jp}

\altaffiltext{1}{Based on data collected at the Subaru Telescope, which is
operated by the National Astronomical Observatory of Japan.}

\altaffiltext{2}{Based on observations obtained at the W. M. Keck Observatory, 
which is operated by the University of California and the California Institute of 
Technology. Keck time has been granted by NOAO and NASA.}

\altaffiltext{3}{Graduate School of Science and Technology, Kobe University, 
1-1 Rokkodai, Nada, Kobe 657-8501, Japan}

\altaffiltext{4}{Okayama Astrophysical Observatory, National Astronomical Observatory,
Kamogata, Asakuchi, Okayama 719-0232, Japan}

\altaffiltext{5}{Department of Physics \& Astronomy, San Francisco State University, 
San Francisco, CA  94132; fischer@stars.sfsu.edu}

\altaffiltext{6}{Center of Excellence in Information Systems, Tennessee State University, 
330 10th Avenue North, Nashville, TN 37203; Also Senior Research Associate, Department of 
Physics and Astronomy, Vanderbilt University, Nashville, TN 37235}

\altaffiltext{7}{UCO/Lick Observatory, University of California at Santa Cruz, 
Santa Cruz, CA 95064}

\altaffiltext{8}{Department of Terrestrial Magnetism, Carnegie Institute of 
Washington DC, 5241 Broad Branch Rd. NW, Washington DC, USA 20015-1305}

\altaffiltext{9}{Department of Astronomy, University of California, Berkeley, CA USA 94720}

\altaffiltext{10}{Tokyo Institute of Technology, Ookayama, Meguro-ku, Tokyo 152-8551, 
Japan; and University of California Observatories, Lick Observatory, University of 
California, Santa Cruz, CA 95064}

\altaffiltext{11}{Space Telescope Science Institute, 3700 San Martin Dr., Baltimore, MD 21218}

\altaffiltext{12}{Fairborn Observatory, HC2 Box 256, Patagonia, AZ  85624}

\altaffiltext{13}{Department of Astronomy, Pontificia Universidad Catolica, Avenida
Vicu\~na Mackenna 4860, Casilla 306 Santiago 200, Chile}

\begin{abstract}
Doppler measurements from Subaru and Keck have revealed radial velocity 
variations in the $V = 8.15$, G0IV star HD 149026 consistent with a 
Saturn-mass planet in a 2.8766 day orbit.  Photometric observations at 
Fairborn Observatory have detected three complete transit events with 
depths of 0.003 mag at the predicted times of conjunction.  HD~149026 is now the 
second brightest star with a transiting extrasolar planet.  The mass of the star, based 
on interpolation of stellar evolutionary models, is $1.3 \pm 0.1$ \msun; 
together with the Doppler amplitude, $K_1$ = 43.3 \ms, we derive a planet 
mass, \msini = 0.36 \mjup, and orbital radius of 0.042 AU.  HD~149026 is 
chromospherically inactive and metal-rich with spectroscopically derived 
[Fe/H] = +0.36, \teff = 6147~K, \logg = 4.26 and \vsini = 6.0 \ks.  Based 
on \teff\ and the stellar luminosity of 2.72 \lsun, we derive a stellar 
radius of 1.45 \rsun.   Modeling of the three photometric transits provides 
an orbital inclination of $85.3 \pm 1.0$ degrees and (including the uncertainty
in the stellar radius) a planet radius of $0.725 \pm 0.05$ \rjup.
Models for this planet mass and 
radius suggest the presence of a $\sim 67$ \mearth\ core composed of elements 
heavier than hydrogen and helium.  This substantial planet core would be 
difficult to construct by gravitational instability.
\end{abstract}


\section{Introduction}

Ongoing precise Doppler surveys by US and European teams are observing
about 2000 of the closest and brightest main-sequence stars, and have
detected more than 150 Jupiter-like extrasolar planets (Marcy \etal
2004, Mayor \etal 2004).  These planets show a wide variety in
their masses and orbital characteristics. Among them, 25 planets with
\msini $> 0.2$ \mjup\ reside in orbits close to the central stars with
$P < 14$ days.  Such short-period giant planets represented by 51 Peg
b (Mayor \& Queloz 1995) are called ``hot Jupiters''.  Recently, ``hot
Neptunes'' with $M\sin i=10-20 M_{\oplus}$ and $P < 14$ days have also
been found (Butler \etal 2004, McArthur \etal 2004, Santos \etal
2004a, Vogt \etal 2005).  The very highest Doppler precisions, $\sim 1$ \ms, are capable
of detecting planets down to about $\sim$ 5 \mearth\ in short period orbits
(Vogt \etal 2005, Rivera \etal 2005).

Short-period planets have provided deep insights into our understanding of planet 
formation, interior structure, atmospheres, and orbital 
evolution of extrasolar planets.  In particular, the detection of a hot 
Jupiter transiting HD 209458 (Henry \etal 2000a, Charbonneau \etal 2000) 
has provided key information about the planet radius, density, and atmospheric 
constituents.  The quality of this information is due in large part to the 
intrinsic brightness of the star ($V=7.65$) which permitted photometric 
precision of $10^{-4}$ using the Hubble Space Telescope (Brown \etal 2001, Wittenmyer \etal 2005).

Atmospheric models predict different radii for planets with and without 
heavy element cores. One mystery regarding HD~209458b is that the planet radius is 
20\% larger than that predicted, even by models without cores. 
This observation has fueled controversy regarding the physical cause 
of the expansion of the radius of HD~209458b. Burrows, Sudarsky, \& Hubbard (2003) 
suggest that current theory, including the effects of stellar heating on 
the atmosphere of the planet but no additional dissipative effects, is consistent
with the radius within theoretical and observational error bars. 
Guillot \& Showman (2002), on the other hand, have calculated  
that the stellar radiative input of energy generates gas flows 
from the heated side to the cool side of the synchronously rotating
planet. Dissipation of a small fraction of this kinetic energy in the 
deeper layers of the planet can result in heating and expansion.
However it is not clear why this same mechanism does not inflate the radii
of other short-period, transiting planets with very similar properties, 
such as TrES--1 (Alonso \etal 2004). 

Tidal dissipation, driven by ongoing eccentricity damping has also been 
suggested as a source of internal heating in HD~209458b 
(Bodenheimer, Lin \& Mardling 2001, Baraffe \etal 2003, 
Bodenheimer, Laughlin \& Lin 2003). However, observations of 
the timing and duration of the secondary eclipse (Deming \etal 2005) and 
the precise radial velocities themselves 
(Laughlin \etal 2005) constrain the orbital eccentricity to be low, 
$e < 0.02$, making eccentricity damping an unlikely source of significant
internal heat for the planet.

One other source for tidal heating has recently been suggested: tidal
dissipation driven by gradual co-planarization between the stellar
equatorial plane and the orbital plane of the planet.  A 4\arcdeg
misalignment between the stellar equatorial plane and the orbital
plane of HD~209458b was determined by modeling the Rossiter-McLaughlin
effect (Winn \etal 2005).  Since the timescale
for co-planarization exceeds the age of the star, this process could
still be driving tidal dissipation in the planet.  It would be useful
to measure this effect in host stars of other transiting planets to
determine whether spin-orbit misalignment is uniquely associated with
HD~209458.

Because short-period planets can substantially advance our understanding 
of planet structure and atmospheres, the N2K consortium (Fischer \etal 2005a) 
was established to carry out a distributed Doppler survey of FGK stars ($7.5 < V <10.5$).  
As telescope time is allocated, an optimal set of stars is selected and 
observed over three consecutive nights. Our Monte Carlo tests show that these 
three observations will flag hot Jupiter candidates with orbital periods 
between 1.2 and 14 days. The N2K project draws from a database of 
$\sim 14000$ stars closer than 100 pc and brighter than $V = 10$.
Metallicities for all stars in the database were established using a 
broadband color calibration (Ammons \etal 2005). Low resolution spectroscopic 
follow-up (Robinson \etal 2005) was obtained for stars
with high metallicity estimates.  High metallicity 
stars are preferred for this program because of the 
well-established planet-metallicity correlation 
(Gonzalez 1997, 1998, 1999, Gonzalez \etal 2001, Fuhrmann \etal 1997, 1998, 
Santos \etal 2001, 2003, 2004b, Reid 2002) which 
shows that stars with [Fe/H]$ > 0.2$ have more than three times as many gas giant 
planets (Fischer \& Valenti 2005).  Ida and Lin (2004) propose that the 
planet-metallicity correlation is quantitatively accounted for 
by a core accretion scenario for formation of gas giant planets. 
Their models also predict the presence of hot Neptunes, 
albeit only around metal-rich stars.

The N2K consortium has screened $\sim750$ stars in the past 10 months using the Keck, 
Magellan and Subaru telescopes.  Three non-transiting close-in gas giant planets have been detected
(HD~88133, Fischer \etal 2005a; HD~149143, HD~109749, Fischer \etal 2005b) and 
follow-up observations are being made for about 40 additional planet candidates. In 
this paper, we report on the first detection flagged with Subaru observations:
a ``hot Saturn'' that has been observed to transit the G0 IV star HD~149026.

\section{The Subaru N2K Survey}
The Subaru component of the N2K survey employs the High Dispersion
Spectrograph (HDS) on the 8.2 m Subaru Telescope (Noguchi \etal 2002).
To provide a fiducial wavelength reference for precise radial velocity (RV) measurements,
we use an iodine absorption cell, which is installed just behind the entrance
slit of the spectrograph (Kambe \etal 2002; Sato \etal 2002).
We adopt the setup of StdI2b which simultaneously covers a wavelength region
of 3500--6100 \AA \ by a mosaic of two CCDs and
the slit width of 0$^{\prime\prime}$.8 giving a reciprocal
resolution ($\lambda/\Delta\lambda$) of 55000. Using this setup,
we can obtain signal-to-noise ratio of $S/N\sim150$ pixel$^{-1}$ at
5500 \AA \ for our typical science targets, $V\sim8.5$ stars, for
an exposure time of about 60 seconds. Such S/N and wavelength resolution
enable us to achieve a Doppler precision of 4--5 \ms.

The set of N2K runs at Subaru were carried out on July 19--21 and August 23
in 2004. The first 3 nights were devoted to identify stars showing
short-period RV variations consistent with a hot Jupiter, and the 4th
night was for follow-up observation to eliminate spectroscopic binaries
and to extend the observational baseline for previously observed stars. 
On the first observation, we checked for
emission in the Ca H\&K lines, and analyzed all spectra with a 
spectral synthesis modeling pipeline to determine 
metallicity, \teff, \logg\ and \vsini\ with uncertainties of 0.05 dex, 40K, 
0.05 dex and 0.5 \ks\ respectively, as discussed in Valenti \& Fischer (2005).  
Spectral modeling also flags double-lined spectroscopic binaries (SB2s)
by virtue of an extremely poor \chisq\ fit to the observation.  Visual 
inspection of the poorly-modeled spectra then confirmed the presence 
of a second set of spectral lines. 
As a result of this screening, 9 of 125 stars were identified 
as SB2s or rapid rotators and were dropped before a second observation 
was obtained.  All information regarding every star observed, 
including RV measurements, information from spectral synthesis modeling,
information about the presence of stellar companions and chromospheric 
activity measurements will appear in an N2K catalog (Fischer \etal 2005c).

During the Subaru runs described above,
we obtained 3--4 Doppler measurements for 116 stars. Typical
instrumental precision of the radial velocities is found to be
3--4 \ms\, with the standard
Doppler pipeline developed by Butler \etal (1996).  With this Doppler precision,
stars displaying RMS scatter between 20--50 \ms\ are likely planet candidates.  
HD~149026 was one of six stars that showed an interesting initial RMS scatter 
(37 \ms over 3 nights in July 2004).  Subsequent observations at Keck in 
February and April 2005 confirmed a planetary orbit and provided 
ephemeris times for photometric follow-up. 


\section{HD 149026}

HD 149026 is a G0 IV star with $V=8.15$ and $B - V = 0.611$. The \hipp\ parallax 
(ESA 1997) of $12.68~ mas$ places the star at a distance of 78.9 pc with an absolute
visual magnitude, $M_V = 3.66$.  Our spectroscopic analysis yields \teff $= 6147 \pm 50$ K,
\fe $= 0.36 \pm 0.05$, \logg $= 4.26 \pm 0.07$ and \vsini $= 6.0 \pm 0.5$ \ks. 
The metallicity prediction from the broadband calibration of Ammons \etal (2005) was $0.36 \pm 0.1$, 
matching the spectroscopic analysis.  
From the bolometric luminosity and our spectroscopic \teff, we derive 
a stellar radius of $1.45 \pm 0.1$ \rsun. The Girardi and Yale evolutionary 
tracks (DeMarque \etal 2004, Girardi \etal 2000) provided identical stellar mass 
estimates of $1.3 \pm 0.1$ \msun.  Although the calibration to measure chromospheric 
activity has not yet been completed for the HDS spectra, no emission was observed in the cores 
of the Ca H\&K lines relative to the NSO solar spectrum (Wallace \etal 1993) 
shown in Figure 1.  The known stellar parameters are summarized in Table 1.

The observation dates, radial velocities and instrumental uncertainties for HD~149026 are listed
in Table 2. The first four observations were made at Subaru; an offset of 
$-$5.4 \ms\ was applied to the Subaru radial velocities in order to minimize \chisq\ when 
fitting a Keplerian model to the combined Subaru and Keck velocities. Amazingly, 4 of 7 
Keck radial velocities were serendipitously obtained in transit.  
Since the transit window is 3 hours long and the orbital period is 2.88 days, 
there is only a $4.3\%$ probability that a randomly chosen observation time will 
occur during transit. The probability of making subsequent in-transit observations 
is not an independent draw, however, since stars are normally observed near the 
meridian. Because the orbital period is close to three days, observations obtained 
on three consecutive nights have an enhanced probability of catching two points 
in transit. Furthermore, two of our in-transit observations were made about an 
hour apart during a single transit.

Our best fit orbital parameters are listed in Table 3 and the Keplerian fit 
is overplotted on the phased radial velocity data in Figure 2. The filled circles 
represent Subaru RV measurements and the open diamonds represent Keck data. The four 
Keck RV's serendipitously taken during transit were removed for purposes of fitting 
the Keplerian model and are overplotted as triangles on Figure 2. 
By fitting to the radial velocity data, we obtain a period of
$P = 2.8766 \pm 0.001$ day. With a fixed circular orbit 
we measure a velocity amplitude of 43.3 \ms.  The RMS RV scatter to this fit is 
3.8 \ms and \chisq = 1.22.  Uncertainties in the orbital parameters were estimated 
by running a bootstrap Monte Carlo with 100 trials.  With the stellar mass of 1.3 \msun, 
we derive \msini = 0.36 \mjup , and $a_{\rm rel} = 0.042$ AU. 

The four radial velocities of HD~149026 obtained during transit exhibit the Rossiter-McLaughlin
(RM) effect, a deviation from Keplerian velocities which occurs because a transiting 
planet occults first the approaching limb of the rotating star and then 
the receding limb of the star. If the orbital plane is coplanar with the stellar 
equatorial plane, then a symmetrical deviation occurs about mid-transit.
The observation of this effect in the Doppler velocities provides an 
unambiguous and independent confirmation of the photometric 
planet transit. The amplitude of this effect in HD~149026 can be estimated by scaling the 
effect in HD~209458 which we estimate at $\Delta v = 45$ \ms.
The photometric transit depth is only 0.003 mag in HD~149026, 
compared to 0.019 mag in HD~209458, however the rotational velocity is somewhat higher
(6 \ks\ compared to 4 \ks). So, we expect a maximum RM amplitude of about
$(3/19)*(6/4)*(45)=11$ \ms, consistent with our four in-transit observations.
If the radius of HD~149026b were similar to HD~209458b, the departure from a 
Keplerian model would have precluded a quick detection. 
Asymmetry in the RM velocities arises from non-coplanar orientations between 
the stellar equatorial plane and the orbital plane.  While the in-transit RVs 
plotted in Figure 2 seem to show an asymmetry, fits that include the constraints provided by photometry 
indicate that this asymmetry is only marginally significant. 

The stellar radius is a critical parameter since only the ratio of the planet radius
to the stellar radius is well-determined by the transit depth. We determine the stellar 
radius from the \hipp -based stellar luminosity, $2.72 \pm 0.5$ \lsun, and our 
spectroscopically-derived \teff, $6147 \pm 50$K:

\begin{equation}
{R \over \rsun} = \sqrt{L \over \lsun} \ \Big( {\tsun \over \teff}\Big)^2 = 1.45 \pm 0.1
\end{equation}

Uncertainty in the stellar radius is dominated by the uncertainty in 
the stellar luminosity.  The limiting factor for precision of the stellar 
luminosity is the \hipp\ parallax precision of 1 $mas$.  Higher 
precision astrometry for HD~149026 could substantially improve the precision in the 
stellar radius. For example, the Space Interferometry Mission (SIM) should achieve 
better than 10 microarcsecond single measurement precision (Shao 2004), 
significantly reducing the uncertainty in the stellar luminosity and 
improving the measurements of stellar and transiting 
planet radii by a factor of six.

\section{Photometric Results for HD 149026}

We began photometric observations of HD~149026 with the T11 0.8 m automatic
photometric telescope (APT) at Fairborn Observatory after the star was recognized
as a short-period radial velocity variable.  Precision photometric
measurements are an important complement to Doppler observations and can
help to establish whether the radial velocity variations are caused by
stellar magnetic activity or planetary-reflex motion (Henry \etal 2000b).
For example, Queloz \etal (2001) and Paulson \etal (2004) have shown that 
photospheric features were the source of periodic radial velocity variations 
in the young G0~V star HD~166435 and in several Hyades dwarfs.  In addition, 
photometric observations can detect possible transits of the planetary 
companions and so allow the determination of their radii and true masses
(e.g., Henry \etal 2000a).

The T11 APT is equipped with a two-channel precision photometer employing two
EMI 9124QB bi-alkali photomultiplier tubes to make simultaneous measurements
in the Str\"omgren $b$ and $y$ passbands.  This telescope and its photometer
are essentially identical to the T8 0.8 m APT and photometer described in
Henry (1999).  The APT measures the difference in brightness between a program
star and a nearby constant comparison star with a typical precision of 0.0015
mag for bright stars ($V < 8.0$).  For HD~149026, we used HD~149504
($V$ = 6.59, $B-V$ = 0.44, F5) as our primary comparison star, which was
shown to be constant to 0.002 mag or better by comparison with two additional
comparison stars (HD~146135, $V$ = 7.08, $B-V$ = 0.36, F2;  HD~151087,
$V$ = 6.02, $B-V$ = 0.32, F2.5~III-IV).  We reduced our Str\"omgren $b$ and
$y$ differential magnitudes with nightly extinction coefficients and
transformed them to the Str\"omgren system with yearly mean transformation
coefficients.  Further information on the telescope, photometer, observing
procedures, and data reduction techniques employed with the T11 APT can be
found in Henry (1999) and Eaton \etal (2003).

A typical observation with the APT, termed a group observation, consists of
several measurements of the program star, each of which are bracketed by
measurements of a comparison star or stars.  The group observations typically
take 5 to 15 minutes to complete, depending on the number of comparison
stars in the group and the integration times for each star.  The individual
differential magnitudes determined within each group are then averaged
to create group means, which are usually treated as single observations.
On most nights, our observations of HD~149026 were obtained and reduced in
this manner, resulting in a total of 153 group mean differential observations
between 2005 April 25 and June 15 UT.  

However, on the nights of 2005 May 14 UT (JD 2453504) and June 6 UT (JD 2453527), 
we monitored HD~149026 and its primary comparison star repeatedly for 
approximately 7.5 hours around the time of conjunction predicted by the 
radial velocity observations and obtained 300 individual differential 
measures on each night.  Rather than averaging the resulting individual 
differential magnitudes within each group on these monitoring nights, 
we have retained the individual observations to increase the time resolution 
of our brightness measurements to about 90 sec during the transit search.  
A few of the monitoring observations from each night were discarded as 
isolated outliers, resulting from such factors as telescope vibration, 
poor centering, or moments of poor seeing.  

Additional {\it simultaneous} monitoring observations were acquired on the 
night of 2005 June 9 UT (JD 2453530) with the T8, T10, and T11 APTs; 330 
differential observations were obtained with each telescope.  The T8 and T10 
APTs are functionally identical to the T11 APT (Henry 1999; Eaton \etal 2003).
These June 9 observations from the three APTs were averaged together into a 
single data stream, and a couple of isolated outliers were discarded.  To 
increase the precision of {\it all} of our total of 973 differential 
magnitudes, we averaged the Str\"omgren $b$ and $y$ magnitudes into a single 
$(b+y)/2$ passband.  The resulting 973 $(b+y)/2$ differential magnitudes 
are listed in Table~4.

All except for a few of our 153 group mean $(b+y)/2$ differential magnitudes 
fall outside of the transit window predicted from the radial velocity 
observations.  The standard deviation of these out-of-transit observations 
is 0.0015 mag, indicating that both HD~149026 and its comparison star are 
constant to the limit of precision of the APTs.  A least-squares sine fit of 
those observations phased to the radial velocity period gives an upper limit 
to the semiamplitude of any light variability on that period of only 
0.0004 $\pm$ 0.0002 mag.  Thus, starspots are unlikely to be the cause of 
the velocity periodicity in HD~149026.  A few of the group mean observations, 
acquired just before dawn on the night of 2005 May 11 UT (JD 2453501), fall 
within the early part of the transit window and suggest a slight dimming of 
around 0.003 mag.  This was our first indication of possible transits in 
HD~149026.  The geometric probability of transits in this system is 
approximately 15\%, computed from equation (1) of Seagroves \etal (2003).

Our three nights of monitoring observations around the predicted times of
conjunction are plotted in Figure~3.  The observations on all three nights 
confirm shallow transits in HD~149026.  The Str\"omgren $(b+y)/2$ 
differential magnitudes have been converted to linear intensity units 
relative to the mean out-of-transit light level measured on each night.  
The monitoring observations have been phased with the 2.8766-day orbital
period and the times of mid-transit determined for each event.  The transits 
occurred approximately 0.05 days or 1.2 hours earlier than predicted 
from the spectroscopic orbital solution but within the limits of its 
uncertainty window.  The observed duration of the transits is just 
over 0.04 phase units or approximately 3 hours, while the observed depth is 
only 0.003 mag, significantly less than the 0.007 mag events we were 
expecting from the stellar radius of 1.45 \rsun (Table~1) and an anticipated 
planetary radius of 1.14 \rjup, corresponding a gas giant planet 
with no core (see \S 5, below). 

We modeled the three sets of transit observations separately with the
eclipsing binary lightcurve synthesis software of Bradstreet \& 
Steelman (2002) to determine the planetary radius and orbital inclination
as well as the times of mid transit.  The stellar mass, radius, and effective 
temperature, along with the planetary mass, orbital period, and eccentricity, 
were fixed at the values given in Tables 1 and 3, respectively.  A linear 
limb darkening coefficient of 0.61 was used (van Hamme 1993), appropriate 
for the stellar $T_{\rm eff}$ and $\log g$ values from Table 1 and the  
Stro\"mgren $(b+y)/2$ bandpass of the observations.  No attempt was made to 
model the $b$ and $y$ lightcurves separately since no color change was 
detectable during the transit and also because the light curves in the 
individual bandpasses exhibited more scatter.  Our best-fit time of
mid-transit, planetary radius, and orbital inclination, as well as the rms 
of each solution, are listed in Table~5.  The weighted means of the
planetary radius and the inclination are $0.725 \pm 0.05$ \rjup\ and
$85.3 \pm 1.0\arcdeg$, respectively, where the third solution was
given twice the weight of the others because of its significantly smaller 
RMS scatter and the radius is relative to Jupiter's equatorial radius, 
\rjup\ = $7.15\times 10^9$ cm.  The error 
bars include the effects of the uncertainties in the assumed parameters, 
the most significant of which is the uncertainty in the stellar radius.  
Our best-fit models for each transit are shown by the solid curves 
in the three panels of Figure~3.  The best-fit model of the third transit,
which has the most precise observations, is shown pictorially in Figure~4.  
The ratio of the stellar to planetary radius is approximately $20:1$, resulting in the 
observed shallow transits of only $0.003$ mag.

Because of the relatively large stellar radius and the short planetary 
orbital period, transits would occur for inclinations down to a limit of 
about 81\arcdeg, where the center of the planet would just graze the stellar
limb.  However, grazing or near-grazing events result in much poorer fits
to the photometric observations, as shown in the top panel of Figure~5, 
where we have plotted the reduced $\chi_{\nu}^2$ of the best-fit model of
the third transit for a range of assumed inclinations from 81\arcdeg to 
90\arcdeg.  As shown in the bottom panel of Figure~5, the best-fit planetary
radius must increase as the inclination decreases, to compensate for crossing 
the star farther and farther from the center of the stellar disk.  As the 
transits become partial below about 82\arcdeg, the planetary radius must 
increase dramatically to maintain the observed transit depth.  However, the 
transit {\it duration} for these grazing or near-grazing transits becomes 
much shorter than the observed duration of the transit, resulting in the 
much higher reduced $\chi_{\nu}^2$ values for the best fits seen in the top 
panel of Figure~5.  Thus, the observations are clearly incompatible with 
grazing or near-grazing transits of a larger planet.

\section{Comparison with Model Radii}

Both the mass and the radius of HD 149026 are considerably smaller than those
of the other known transiting extrasolar planets.
Having estimated the planetary radius, $R=0.725 \pm 0.05~R_{\rm Jup}$ and
mass, $M=0.36 \pm 0.04~M_{\rm Jup}$ by fitting the radial velocity and transit
photometry data, we are in a position to compare the results with theoretical
evolutionary models. To do this, we use the results of Bodenheimer \etal (2003),
who computed sequences of models for isolated planets ranging in mass from
0.11 $M_{\rm Jup}$ to 3.0 $M_{\rm Jup}$. Separate sequences were computed for
models that contained and did not contain solid, 
constant density ($\rho=5.5~{\rm g~cm}^{-3}$) cores.
Adopting the stellar and orbital parameters listed in Tables 1 and 3, and 
assuming a Bond albedo $A=0.3$, we estimate an effective 
temperature for the planet of $T_{\rm eff}=1540~{\rm K}$. 
This value is uncertain to 10\% because the albedo is not
known. The planet falls on the borderline between
Sudarsky, Burrows, \& Pinto's (2000) Class IV (typical 
albedo 0.03) and Class V (typical albedo 0.55).  The 
corresponding uncertainty in the theoretical radius is
roughly 5\%. Under these conditions, the models of Bodenheimer \etal (2003) 
predict a planetary radius of $R=1.14~R_{\rm Jup}$ 
for an object of purely solar composition, and 
$R=0.97~R_{\rm Jup}$  for a planet with a $20~M_\oplus$ constant density
($5.5~{\rm g~cm}^{-3}$) core. Clearly, in order to 
explain the radius of HD 149026b, a substantial enrichment in heavy elements
above solar composition is required.  The mean density of the planet, 1.17 g cm$^{-3}$,
is 1.7 times that of Saturn, which itself has roughly 25\% heavy elements by mass.
On the other hand, the planet is not composed entirely of water ice or
olivine, or else the radius would be 0.43 or 0.28 \rjup, respectively 
(Guillot \etal 1996, Guillot 2005). 

We have therefore computed new sequences of contracting and cooling planetary
models of mass 0.36 \mjup with the code described
in Bodenheimer et al (2003), under the assumption that that planet is made up
of a core of uniform density composed entirely of elements heavier than helium, and an envelope
composed primarily of fluid with solar composition. It is of course
possible that some of the heavy elements are not in the core but are dissolved
in the envelope; however the effects on the computed mean density will be 
similar (Guillot 2005).  The calculations were started at a radius of 2--3 times the present
value and were run up to an age of 2 Gyr, the estimated age of  HD 149026.
The planet is heated at the surface to 1500 K during the entire evolution.
The surface boundary conditions are those for a gray radiative photosphere
with a Rosseland mean opacity. The deduced radii are therefore approximate
but nevertheless in reasonable agreement with more detailed calculations with
non-gray photospheres (Chabrier \& Baraffe 2000, Burrows \etal 2003, Burrows \etal 2004,
Chabrier \etal 2004). The envelope opacities in the
outer radiative zone are purely molecular (R. Freedman 2003, private 
communication) and do not take into account the possible effects of clouds. 
The equation of state in the envelope is calculated from the tables of 
Saumon, Chabrier, \& van Horn (1995).

The results are surprising.  Assuming a core density
$\rho_c=10.5~{\rm g~cm^{-3}}$, the estimated mean core density of
Saturn (Marley 1999), we find for ($M=0.36\, M_{\rm Jup}$, and
$R=0.725\, R_{\rm Jup}$) a solid core mass of $67\, M_{\oplus}$. If
$\rho_c=5.5~{\rm g~cm^{-3}}$, the required core mass rises to 78
$M_{\oplus}$. For these two assumed core densities, which probably bracket the
actual mean core density that would be obtained with a more detailed
compressible equation of state for the core, the results for the
derived core mass as a function of observed radius are given in Table 6 
for a mass of 0.36 M$_{\rm Jup}$. The interpolated core mass lies between 
entries (A) and (B) in Table 6 for $\rho_c = 10.5~{\rm g~cm^{-3}}$, and between 
entries (C) and (D) for $\rho_c = 5.5~{\rm g~cm^{-3}}$.

For both cases of assumed central density, the core comprises the majority of the
planetary mass, indicating that despite its mass and orbital period,
the planet is qualitatively more similar in structure to Neptune than
it is to Saturn or Jupiter. Furthermore, the conclusion that the core
is large seems secure. Even if the observed radius is increased by 2 sigma 
to 0.825 \rjup, the core mass turns out to be 50.5 \mearth in the 
case of $\rho_c$=10.5 and 58.7 \mearth in the case of $\rho_c$=5.5 g cm$^{-3}$.

Note that the radii computed here do not include an effect pointed out by
Burrows et al. (2003), namely that the radius deduced from
observations of a transit is larger by about 10\% (for HD 209458b) than 
the theoretical ``photospheric'' radius because of the 
oblique viewing angle. For the lower-mass planet HD 149026b
this effect could be even larger. If the planet radius is
actually smaller than 0.7 \rjup, then clearly the
required enhancement of heavy elements would be even larger
than quoted here.

\section{Discussion}

The N2K consortium was established 
to survey ``the next 2000'' closest and brightest high metallicity stars.  
Here we describe a transiting Saturn-mass planet from the N2K consortium 
with an orbital period of 2.8766 days orbiting the $V=8.15$ star, HD~149026. 
The planet was initially identified with 
Doppler observations from Subaru and confirmed with Keck observations.  
Remarkably, HD~149026 only dims by 0.003 mag during the 
three hour transit of the planet.  The photometric transit depth, together with the 
stellar radius of 1.45 \rsun\ provides a planet radius of only 0.725 \rjup.
Four of seven Keck radial velocities were serendipitously 
obtained during transit.  These four radial velocities exhibit the Rossiter-McLaughlin
effect and provide an independent confirmation of the planetary transit. 

With a radius of only 0.725 \rjup, HD~149026b seems anomalously 
small and emphasizes the diversity of extrasolar planet characteristics.    
Indeed, two of the three transits around bright stars (HD~209458, TrES--1 and HD~149026) are 
``anomalous.''  HD 209458 is too big, HD 149026 is too small; only TrES--1 conformed to 
prior expectations.  These anomalies indicate a wider-than-expected diversity of 
planetary types, and may also point to the need for models that include a wider
variety of physical and atmospheric phenomena. 

The small photometric depth of the transit underscores 
the most novel property of HD 149026b -- its large core. 
The apparent presence of the large core has a number of
potentially interesting ramifications for the 
theory of planet formation.
First, it would be difficult to form this giant planet by 
the gravitational instability mechanism (Boss 2004), 
and by corollary, it likely illustrates a dramatic outcome of 
the core-accretion scenario (e.g. Hubickyj \etal 2004, and 
references therein). Among the giant planets in our solar system, 
Neptune is the hardest to account for within the framework 
of gravitational instability. In that
picture, Neptune began as a $\sim 3 M_{\rm Jup}$ Jeans-unstable 
fragment. As the massive, cool  proto-Neptune underwent Kelvin-Helmholtz 
contraction, its burden of solid grains settled to the 
central regions to form a core. As this was occurring, it is 
suggested (Boss 2003) that a nearby massive star photoevaporated 
the outer layers, leaving, at the end of the day, a 17 $M_{\oplus}$ 
remnant of heavy elements, if a solar-composition is assumed.

In order for this scenario to work for HD 149026b, the 
initial twice-solar metallicity fragment would need to have 
a total mass of order 6 $M_{\rm Jup}$ in order to give 
rise to a 67 $M_{\oplus}$ core, which is a total mass of solid components
in the fragment. For such a large protoplanet, 
the settling time for the solid grains is considerably longer than 
the Kelvin-Helmholtz contraction time for the envelope.  
Furthermore, near-complete photo-evaporation of a 
$6 M_{\rm Jup}$ protoplanet is difficult to accomplish.

However, the large core of HD 149026b also presents difficulties 
for conventional models of core accretion. In the 
core-accretion theory, which was developed in the context 
of the minimum-mass solar nebula, it is difficult to prevent 
runaway gas accretion from occurring onto cores more massive 
than 30 $M_{\oplus}$, even if abundant infalling planetesimals 
are heating the envelope and delaying the Kelvin-Helmholtz contraction 
that is required to let more gas into the planet's Hill sphere. 
The current structure of HD 149026b suggests that it was formed 
in a gas-starved environment, yet presumably enough gas was 
present in the protoplanetary disk to drive migration from
its probable formation region beyond one or two AU from the star 
inward to the current 2.87 day orbit.

We note, however, that the buildup of the large core mass was 
aided by the high metallicity $Z$ of the protoplanetary gas. 
In the standard theory, the isolation mass of a protostellar 
core scales as $M_{\rm iso} \propto (a^2\sigma)^{3/2} M_\ast^{-1/2}$,
where $\sigma$ is the surface density of solids in the disk and $a$ is 
the distance from the star, which has mass $ M_\ast$.  We assume 
that $\sigma$ scales as  $M_\ast Z$.
The final core mass, after further accretion of gas and
solids, is about $\sqrt{2} M_{\rm iso}$ (Pollack et al. 1996). 
For a twice-solar metallicity disk around a 1.3 $M_{\odot}$ star, 
this implies a factor of 3.7 increase in  $M_{\rm iso}$ and $M_{\rm core}$ 
as compared to  their values  in the minimum mass solar nebula. 
But at least a  factor of 3 increase in $\sigma$ above that in
the minimum mass solar nebula is required for the core accretion 
model to produce Solar System giant planets in a time comparable 
to the lifetime of protoplanetary disks.  
As an example we use the result of Hubickyj et al. (2004) for
the formation of Jupiter at 5.2 AU around the Sun with $\sigma = 10$ g
cm$^{-2}$.  They obtain  $M_{\rm iso} = 11.4 M_\oplus$ and 
$M_{\rm core} = 16.2 M_\oplus$. Scaling from that result, a plausible 
$M_{\rm core}$ for HD 149026b forming at 5.2 AU is close to 60 $ M_\oplus$.
Even if no further accretion of solids occurs after the isolation mass
is reached, $M_{\rm core}$ would be 42 $ M_\oplus$.  
Planet formation models with shorter accretion timescales may be key 
to understanding the formation of HD~149026b (c.f., Aliberti \etal 2005).

Several scenarios for forming an Saturn-mass object with a 
$\sim 67 M_{\oplus}$ core can be envisioned. The planet may have 
migrated inward when it was less massive than present, and 
became stranded at the 2:1 resonance with the so-called X-point 
of the protostellar disk (e.g. Shu \etal 1994, see also 
the suggestion in Lin, Bodenheimer, \& Richardson 1996). After 
the planetary core was stranded, gas migrating through the disk 
would climb the magnetic field lines onto the star, whereas 
planetesimals would cross through the X-point and be pushed 
further inward by the disk torque. Some of these planetesimals 
would accrete onto the stranded planet, increasing its 
compositional fraction of heavy elements (Ward 1997).

Alternately, a planet with a large core could be formed through
a giant impact scenario between two isolation-mass embryos.       
Such an object could then either migrate inward via conventional 
Type II migration, or perhaps have been scattered into a highly 
eccentric orbit by interactions with other planets in the system. 
In the case of a scattering history, the orbital plane of the planet 
was likely to have been initially misaligned with the spin axis of the 
star. This scenario may soon find support as more in-transit RVs 
which exhibit the Rossiter-McLaughlin effect on the stellar lines 
are obtained (e.g. Winn et al 2005).

There were two necessary components for this transit detection.  First, the 
initial Doppler reconnaissance at Subaru and Keck detected the low-mass, 
short-period planet and provided accurate transit predictions.  Second, the 
robotic telescopes at Fairborn Observatory were available to obtain baseline 
photometry and to carry out millimag-precision photometric observations at 
the predicted transit times.  This approach, combining a quick-look, Doppler 
survey with targeted high-precision photometry has turned out to be an 
efficient way to detect transiting planets around bright stars;  both 
HD~209458 and HD~149026 were discovered with this strategy.

The small radius of this transiting planet bears upon the 
expected detection rate by widefield photometric transit surveys of bright 
stars. Such a small transit depth would likely be missed with the typical photometric 
precision of 0.01 mag.  Another factor that may negatively impact detection 
rates for wide field transit surveys is the Malmquist bias. 
Because of this effect, subgiants are preferentially 
included in magnitude-limited samples.  The larger stellar radii of 
these stars reduces the transit depth and demands millimag observational precision. 
Of the four stars with short-period planets detected in the N2K program,
all have somewhat extended radii. 
HD~88133 has a radius of 1.93 \rsun\ and a planet mass, \msini = 0.25 \mjup; 
HD~149143 has a radius of 1.49 \rsun\ and a planet mass of 1.32 \mjup; HD~109749 has 
a stellar radius of 1.24 \rsun\ and a planet mass of 0.27 \mjup, and HD~149026 
has a radius of 1.45 \rsun\ and a planet mass of 0.36 \mjup.  While, the transit probabilities 
for all four of these detections exceed the canonical 10\% transit probability because 
of the distended stellar radii, the predicted transit depths are only a 
fraction of the transit depth observed for HD~209458. 

At one time, photometric surveys were predicted to detect hundreds of transiting 
planets per month (Horne 2003).  Instead, transit detection has turned out to 
be harder than anticipated, perhaps because the high-precision regime is 
technically difficult for any technique and because the diversity of 
planet characteristics continues to surprise us.  The combination 
of the Doppler technique, which is sensitive to the gravitational 
effect of the planet, and targeted high-precision photometry 
provide complementary information that dramatically increases our 
understanding of extrasolar planets.

\acknowledgements
We thank Akito Tajitsu for his expertise and support of the Subaru HDS 
observations.  We gratefully acknowledge the dedication and support of the 
Keck Observatory staff, in particular Grant Hill for support with HIRES. We 
thank the NOAO and NASA Telescope assignment committees for generous 
allocations of telescope time. Data presented herein were obtained at the 
W. M. Keck Observatory from telescope time allocated to the National 
Aeronautics and Space Administration through the agency's scientific 
partnership with the California Institute of Technology and the University of 
California.  The Observatory was made possible by the generous financial 
support of the W. M. Keck Foundation.  We thank the Michaelson Science Center 
for travel support through the KDPA program.  DAF is a Cottrell Science Scholar 
of Research Corporation. We acknowledge support from NASA grant (to DAF); 
NASA grant NCC5-511 and NSF grant HRD-9706268 (to GWH); NASA grant NAG5-75005 
(to GWM); NSF grant AST-9988358 and NASA grant NAG5-4445 (to SSV); 
NASA grant NAG5-13285 to PB; and NASA 
grant NNA04CC99A (to GL). GWH also acknowledges timely support by David 
Bradstreet for requested modifications to his lightcurve analysis software 
to fit the unexpectedly shallow transits in HD~149026 and thanks Stephen
Henry for assistance in the preparation of Figure~5. DM is supported by 
FONDAP N. 15010003.  This research has made use of the Simbad database, 
operated at CDS, Strasbourg, France.  The authors extend thanks to those of 
native Hawaiian ancestry on whose sacred mountain of Mauna Kea we are 
privileged to be guests.  Without their generous hospitality, the Subaru and 
Keck observations presented herein would not have been possible.

\clearpage

\begin{figure}
\plotone{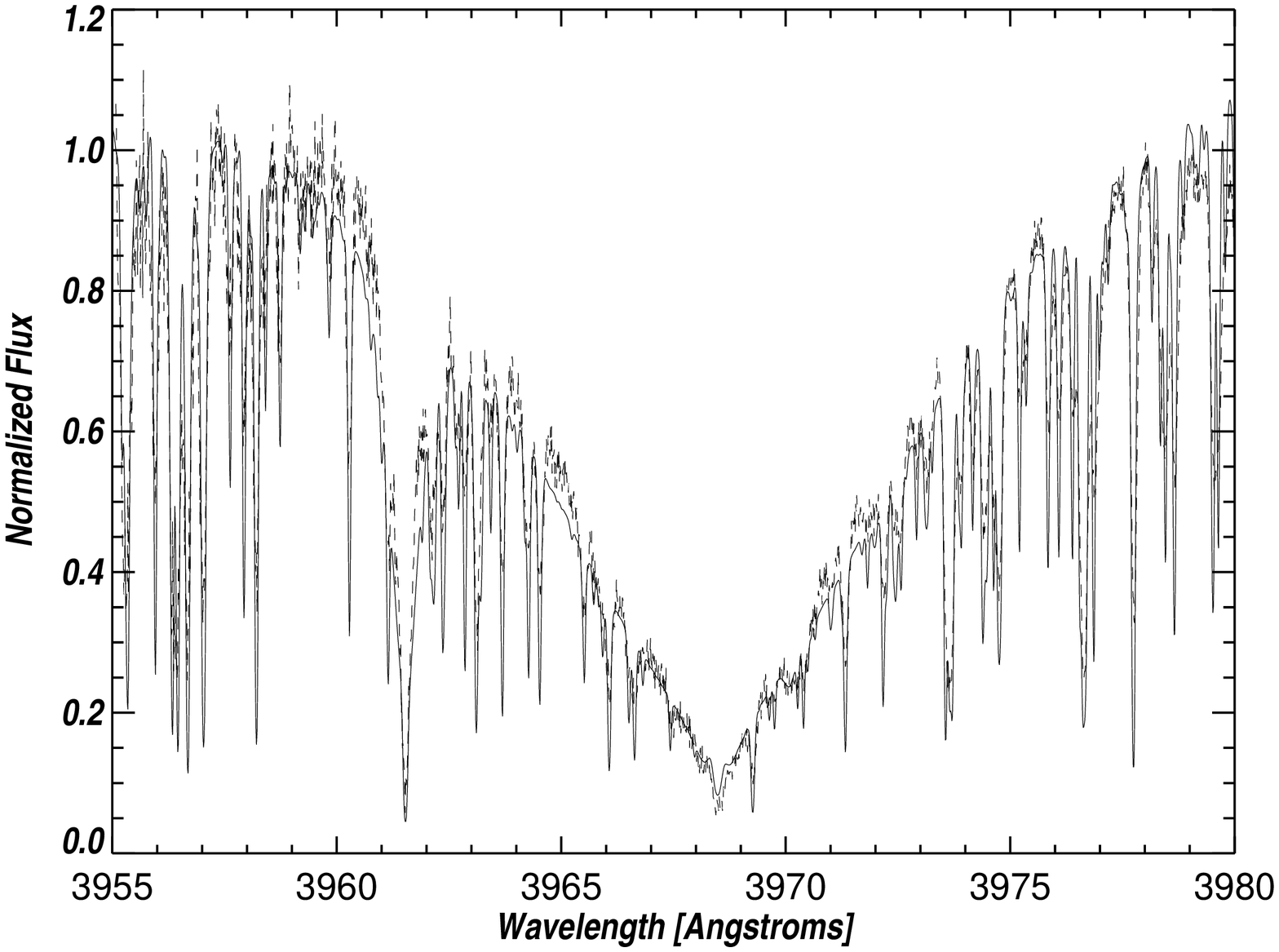}
\figcaption{Ca H line for HD~149026 is plotted as a dashed line with the NSO 
solar spectrum overplotted for comparison. No emission is 
seen in the line core, indicating low chromospheric activity for 
this star }
\label{fig1}
\end{figure}
\clearpage

\begin{figure}
\plotone{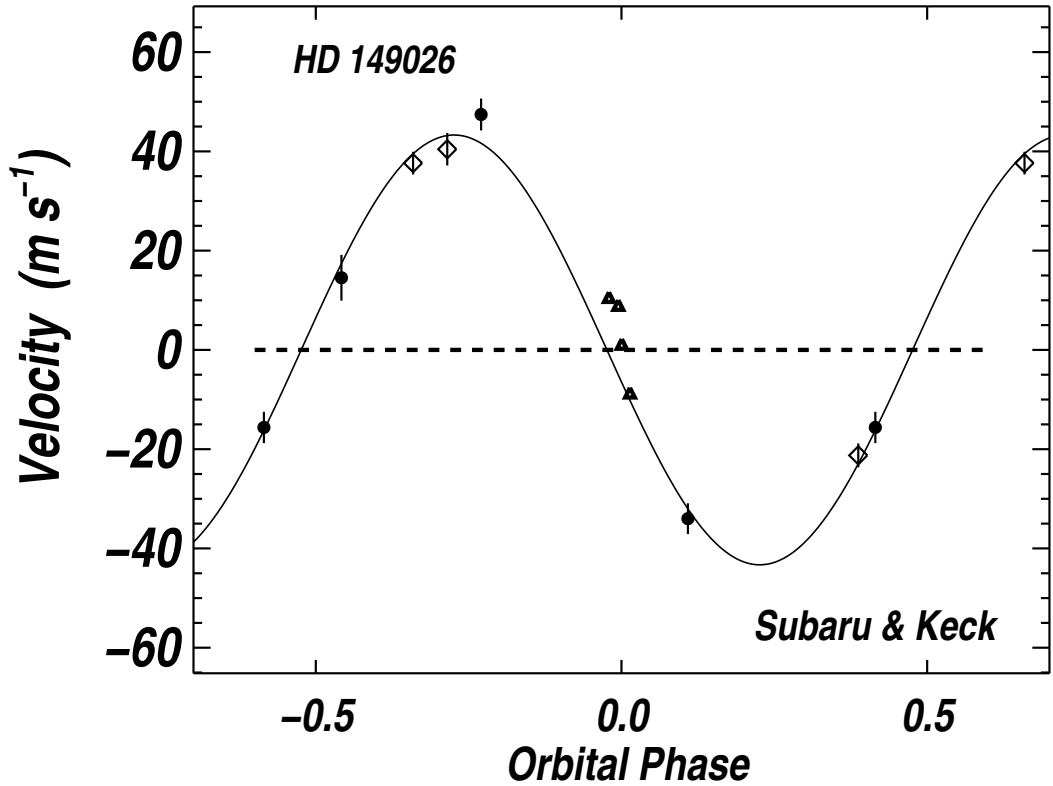}
\figcaption{Phased radial velocities for HD~149026 from 
Subaru (filled circles) and Keck (open diamonds). The 
triangles represent RV measurements made at Keck during transit.
These velocities exhibit the Rossiter-McLaughlin effect 
and were removed when fitting a Keplerian model. While there 
appears to be an asymmetry in the in-transit velocities, fits 
that include the constraint provided by the observed transit midpoint
indicate that this asymmetry is only marginally statistically 
significant.  With an orbital period of 2.8766 day, velocity amplitude
of 43.3 \ms\ and stellar mass of 1.3 \msun, the implied 
planet mass is \msini = 0.36 \mjup\
and the orbital radius is 0.042 AU. }
\label{fig2}
\end{figure}
\clearpage

\begin{figure}[t!]
\figurenum{3}
\epsscale{1.0}
\plotone{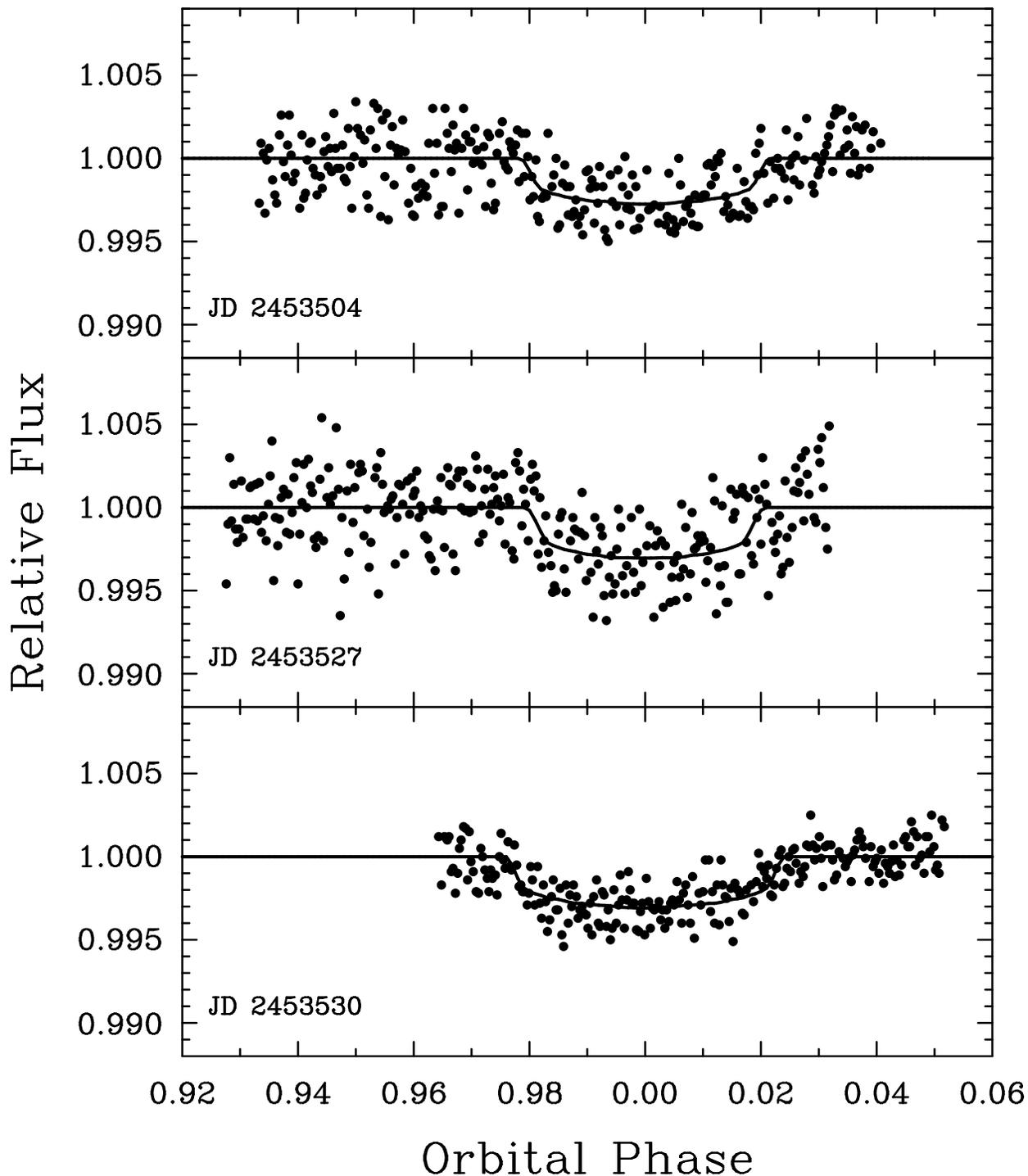}
\figcaption{Photometric transits of HD~149026 in the Str\"omgren $(b+y)/2$ 
passband observed with the T11 APT at Fairborn Observatory on 2005 May 14 UT 
($top$) and 2005 June 6 UT ($middle$) and with the T8, T10, and T11 
APTs on 2005 June 9 UT ($bottom$).  The data from the three telescopes 
for the June 9 event have been averaged together to increase the precision.  
The solid curves represent the best-fit models given in Table~5, which result
in a mean planetary radius of $0.725 R_{\rm Jup}$.  This small planetary 
radius suggests that the planet has a substantial heavy-element core.}
\end{figure}

\begin{figure}[t!]
\figurenum{4}
\epsscale{1.0}
\plotone{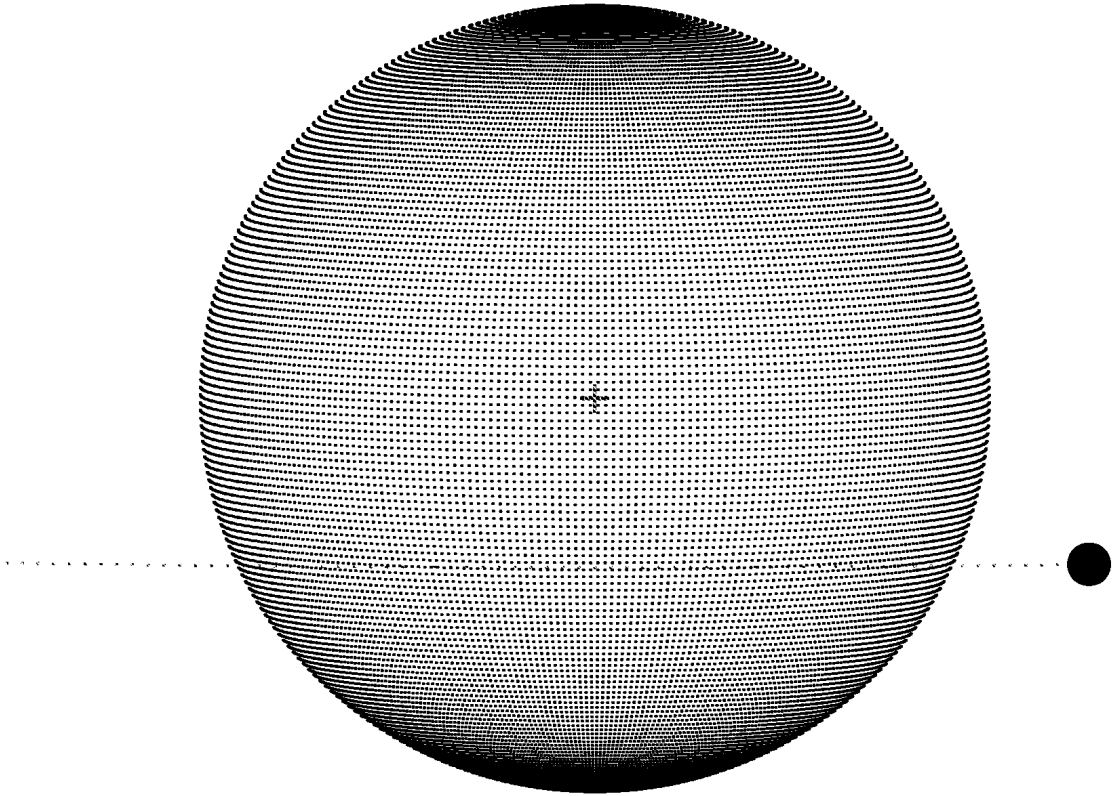}
\figcaption{Depiction of the transit chord for the HD~149026 system based on the third transit
solution.  The ratio of the stellar to planetary radii is approximately 20 to 1, which 
when combined with the appropriate limb-darkening law results in the shallow observed
transits of only 0.003 mag.  It is unknown 
whether or not the planetary orbital plane is actually aligned with the 
stellar equatorial plane (see \S 3).}
\end{figure}

\begin{figure}[t!]
\figurenum{5}
\epsscale{1.0}
\plotone{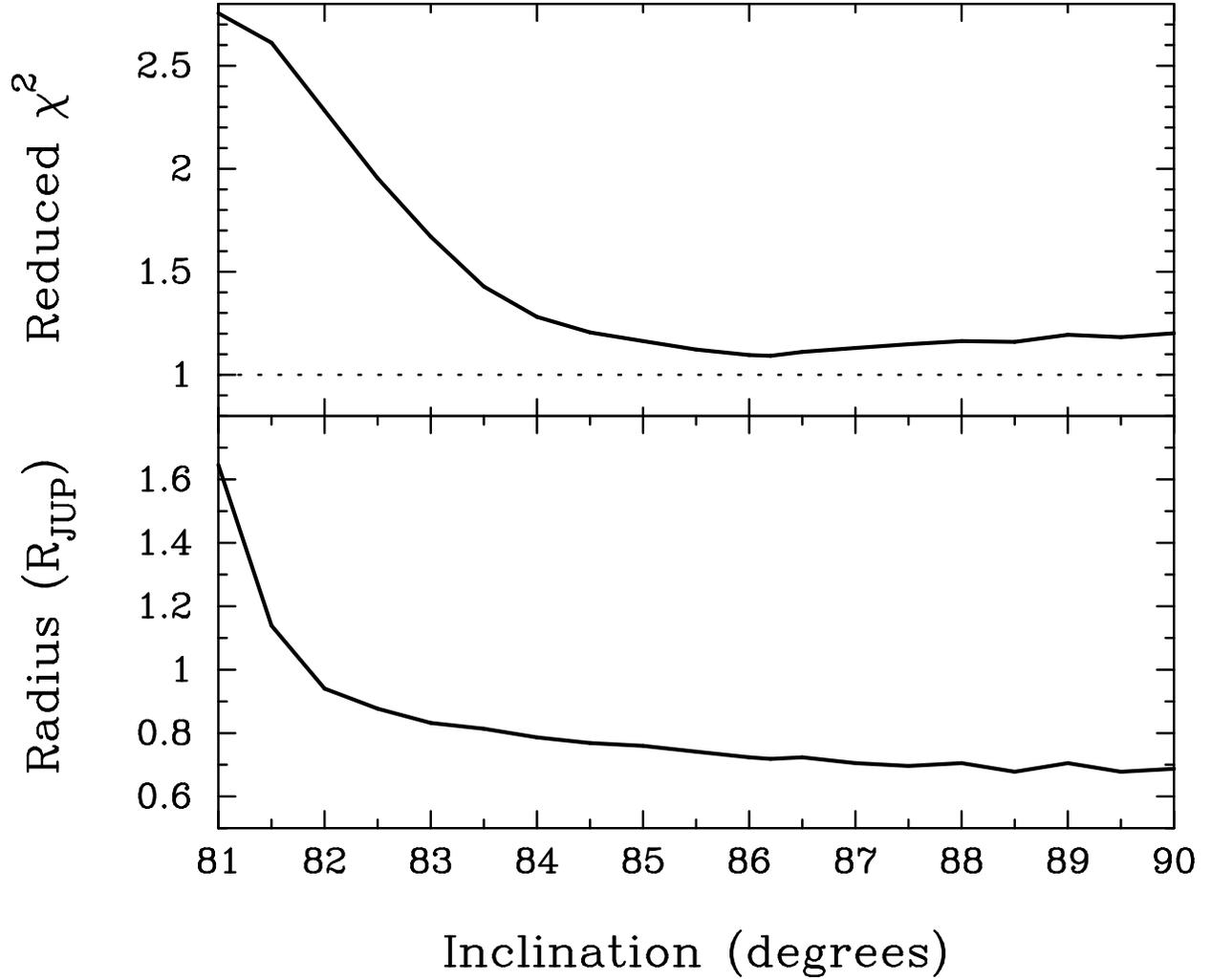}
\figcaption{Reduced $\chi_{\nu}^2$ values ($top$) for the best-fit
planetary radius over a range of assumed inclinations computed for the
2005 June 9 transit.  As the inclination decreases, the best-fit planetary
radius must increase ($bottom$) to compensate.  These results demonstrate
that the observations are incompatible with grazing or near-grazing
transits of a larger planet.}
\end{figure}

\begin{deluxetable}{lll}
\tablenum{1}
\tablecaption{Stellar Parameters for HD 149026}
\tablewidth{0pt}
\tablehead{\colhead{Parameter}  & \colhead{Value} \\
} 
\startdata
$V$                & 8.15          \\
$M_V$              & 3.66          \\
$B-V$              & 0.611         \\
Spectral type      & G0 IV         \\
Distance (pc)      & 78.9 (6.6)    \\
$T_{\rm eff}$ (K)  & 6147 (50)     \\
\logg              & 4.26 (0.07)   \\
${\rm [Fe/H]}$     & 0.36 (0.05)   \\
\vsini (\ks)       & 6.0 (0.5)     \\
$M_{STAR}$ (\msun) & 1.3 (0.1)    \\
$R_{STAR}$ (\rsun) & 1.45 (0.1)    \\
$L_{STAR}$ (\lsun) & 2.72 (0.5)    \\
Age (Gyr)          & 2.0 (0.8)     \\
\enddata                         
\end{deluxetable}                           
\clearpage

\begin{table}
\tablenum{2}
\caption{Radial Velocities for HD~149026}
\begin{tabular}{rrrc}
\tableline
\tableline
       JD  &   RV    & Uncertainties  & Observatory \\
 -2440000  &  (\ms)  &   (\ms)        &             \\
\tableline
   13206.913796  &  -24.10  &   3.15    & Subaru  \\      
   13207.935648  &   38.95  &   3.21    & Subaru  \\
   13208.908681  &  -42.48  &   3.10    & Subaru  \\
   13241.797685  &    6.06  &   4.61    & Subaru  \\
   13427.158623  &    0.76  &   2.08    & Keck\tablenotemark{*}    \\
   13429.113935  &   27.76  &   2.29    & Keck    \\
   13430.079410  &   -0.82  &   2.02    & Keck\tablenotemark{*}     \\
   13430.095787  &   -8.66  &   2.15    & Keck\tablenotemark{*}     \\
   13479.036481  &  -18.49  &   3.36    & Keck\tablenotemark{*}     \\
   13480.110602  &  -31.13  &   2.41    & Keck    \\
   13483.930012  &   30.55  &   3.27    & Keck    \\
\end{tabular}
\tablenotetext{*}{Radial velocities serendipitously obtained during 
transit.} 
\end{table}
\clearpage

\begin{deluxetable}{ll}
\tablenum{3}
\tablecaption{Spectroscopic Orbital Solution for HD 149026b}
\tablewidth{0pt}
\tablehead{\colhead{Parameter}  & \colhead{Value} \\
} 
\startdata
$P$ (days)               &  2.8766 (0.001)      \\
$T_{\rm c}$ (JD)   &  2453317.838 (0.003) \\
Eccentricity             &  0 (fixed)        \\
$K_1$ (\ms)              &  43.3 (1.2)          \\
$a$ (AU)                 &  0.042               \\
$a_1 \sin i$ (AU)        &  1.037e-05          \\
$f_1$(m) (\msun)     &  1.839e-11          \\
$M\sin i$ (\mjup)    &  0.36 (0.03)        \\
${\rm Nobs}$             &  7 (out of transit)     \\
RMS (\ms)                &  3.8              \\
Reduced \chisq           &  1.22              \\
\enddata                        
\end{deluxetable}                          
\clearpage

\begin{deluxetable}{ccc}
\tablenum{4}
\tablewidth{0pt}
\tablecaption{PHOTOMETRIC OBSERVATIONS OF HD~149026}
\tablehead{
\colhead{Hel. Julian Date} & \colhead{$\Delta (b+y)/2$} \\
\colhead{(HJD $-$ 2,400,000)} & \colhead{(mag)}
}
\startdata
53,485.7969 & 1.6056 \\
53,485.8832 & 1.6031 \\
53,486.7939 & 1.6035 \\
53,486.8799 & 1.6036 \\
53,487.7216 & 1.6013 \\
\enddata
\tablecomments{Table 4 is presented in its entirety in the electronic edition
of the Astrophysical Journal.  A portion is shown here for guidance regarding
its form and content.}
\end{deluxetable}

\begin{deluxetable}{ccccccc}
\tablenum{5}
\tablewidth{0pt}
\tablecaption{PHOTOMETRIC SOLUTIONS FOR HD~149026b}
\tablehead{
\colhead{} & \colhead{} & \colhead{Date} & \colhead{$T_{mid}$} & \colhead{Planetary Radius} & \colhead{Inclination} & \colhead{rms} \\
\colhead{Transit} & \colhead{APT} & \colhead{(UT)} & \colhead{(HJD)} & \colhead{($R_{\rm Jup}$)} & \colhead{(degrees)} & \colhead{(mag)}
}
\startdata
      1       &   11    & 2005 May 14 & 2,453,504.865 & 0.705 & 84.6 & 0.0017 \\
      2       &   11    & 2005 June 6 & 2,453,527.864 & 0.759 & 84.1 & 0.0021 \\
      3       & 8,10,11 & 2005 June 9 & 2,453,530.751 & 0.719 & 86.2 & 0.0012 \\
Weighted Mean &         &             &               & 0.725 & 85.3 &        \\
\enddata
\end{deluxetable}

\begin{deluxetable}{lll}
\tablenum{6}
\tablewidth{0pt}
\tablecaption{MODEL RADIUS AND CORE MASS}
\tablehead{
\multicolumn{2}{c}{Radius (\rjup, equatorial)}  & \colhead{Core Mass (\mearth)} \\
\colhead{$\rho_{c} = 10.5~{\rm g~cm^{-3}}$} & \colhead{$\rho_c = 5.5~{\rm g~cm^{-3}}$} & \colhead{ }}
\startdata
      0.594       &  0.662 (C)    & 89.3 \\
      0.681 (A)   &  0.745 (D)    & 74.5 \\
      0.769 (B)   &  0.818        & 60.0 \\
      0.866       &  0.905        & 43.6 \\  
\enddata
\end{deluxetable}

\end{document}